\documentclass{elsart}
\usepackage{natbib}
\begin{document}

\runauthor{}

\begin{frontmatter}

\title{Analytic Derivation of the Quark Distribution Functions Near x=1 by Statistical Methods}
\author{Hai Lin ${}_{}^{1}$}
\address{Department of Physics, Peking University, P.R.China, 100871}
\thanks[Someone]{Email:hailin@mail.phy.pku.edu.cn} 

\begin{abstract}
In this paper, we made a statistical approach on the thermodynamic structure of the nucleon and its quark distributions. 
We assume that the nucleon is a thermodynamic system of quarks and gluons. After we derived the quark density of states 
from the Dirac Equation,  we calculated each quark's chemical potential, associating it with the quark's spin distribution. 
By using the sum rule of the longitudinal momentum, we obtain a function showing the relation between the temperature, radius
and mass of the nucleon. The nucleon's radius is determined by the total energy minimal principle, and thus the temperature 
is also known. Finally, we deduced the quark distribution functions near x=1, and the results are in agreement with
the experimental data.     
\end{abstract}
\end{frontmatter}

\section{Introduction}   
It has been justified by several authors in the literature that the nucleon, which confines many sea quarks and gluons,
could be considered as a thermodynamic system with temperature $T$ in a spherical volume with the radius $R$ [1], [2], 
[3], [4], [5]. 
The statistical method based on the grand canonical ensemble formalism is then introduced to study the thermodynamic
 properties of the nucleon [6].
 After R. S. Bhalerao found a relation between the quark distibution functions $q(x)$ in the infinite momentum frame and 
its density of states $\rho (E)$ in the nucleon rest frame [3],
\begin{equation}
q(x)=\frac{{M}_{}^{2}x}{2} {\int}_{Mx/2}^{M/2} { g f(\mu) \rho (E) \frac {dE}{{E}_{}^{2}} },   
\end{equation}
where $M$ is the nucleon mass, $g$ is the quark degeneracy factor and $f(\mu)$ is the Fermi distribution function,
the statistical methods can be used to calculate various distribution 
and structure functions of the nucleon [7]. \\
However, in order to settle down all the parameters in eq (1), 
we have to find out the quark density of states and their chemical potentials at first.
 In this paper, we deduce the density of states by Dirac Equation and obtain the chemical
potentials by inputting $\Delta u$, $\Delta d$, $\Delta s$. Our method differs from Bhalerao's model [7]
 in that: \\
1. We use a new density of states derived from D.E.; \\
2. We determine the nucleon's radius by the T-R-M function and the total energy minimal principle; \\
3. We use analytic methods, and our results are only about the $x \approx 1$ region. Futher results
may able to be extented to small x region if we use numerical methods. 

\section {Dirac Equation and Quark Density of States} 
In our recently proposed Quark Gluon Coupling Model [6], we describe the nucleon as confining the 
quarks and gluons in a spherical volume with the radius of $R$. 
The quarks interact through the exchange of gluons, 
and the gluon couples to the conserved quark-current. The Lagrangian density for this field is 
\begin{equation}
L=\overline{\psi }\left[ \gamma ^{\mu }\left( i\partial _{\mu}-
{g}_c A_{\mu }\right) -m_{q}\right] \psi -\frac{1}{4}F_{\mu \nu }F^{\mu \nu},  
\end{equation}
where  $A_{\mu}$ is the gluon vector field, ${g}_c$ is the coupling constance, $m_{q}$ is the quark mass and
 $F_{\mu \nu }= \partial _{\mu }A_{\nu }-\partial _{\nu }A_{\mu }$.
If we replace the vector field operators by their expectation values, that is, $
A_{\mu  }\rightarrow \delta _{\mu 0}A_{0}$,
the Lagrange's equations yield the the linear Dirac Equation, 
\begin{equation}
\left[ i\gamma ^{\mu }\partial _{\mu }-g_c\gamma ^{0}A_{0}-m_{q} \right] \psi =0.
\end{equation}
The normalized quark wave function within a spherical volume with the radius of $R$ is
\begin{equation}
\psi({\vec{r}},t) = {N} e^{-i E t}
\times \left( \begin{array}{c} { \sqrt{\frac{E-{g}_c{A}_0+{m}_q}{E}}
  j_0( \sqrt{{(E-{g}_c{A}_0)}_{}^{2}-{m}_{q}^{2}} r)} \\
 i{ \sqrt{\frac{E-{g}_c{A}_0-{m}_q}{E}} j_1( \sqrt{{(E-{g}_c{A}_0)}_{}^{2}-{m}_{q}^{2}} r))
  }\end{array} \right)
{{{\chi}_q}\over{\sqrt{4 \pi}}} ,
\end{equation}
where $r=|{\vec{r}}|$,  ${\chi}_q$ is the quark spinor and $N$ is the normalization constant [8] and 
$E$ is the quark energy eigenvalue.
We note that $\overline{\psi }\psi $ should vanish
at the boundary in the relativistic theory [9], $
\left. \overline{\psi }\psi \right| _{r=R}=0.$ 
This yields the energy states [6]:
\begin{equation}
{E}_{n}^{} \approx g_c{A}_0+\sqrt{{m}_{q}^{2}+{(\frac{n\pi}{R})}_{}^{2}} ,
\end{equation} 
 where $n=1,2,...,$  and therefore the quark density of states is
\begin{equation}
{\rho}(E)=\frac{R(E-g_c{A}_0)} {\pi \sqrt{{(E-g_c{A}_0)}_{}^{2}-{m}_{q}^{2}}}.
\end{equation}
If we use the assumption in the parton model that $m_q \rightarrow 0$, the density of states will be reduced to $
 {\rho}(E)=\frac{R}{\pi}$. We will use this density of states in all of the following sections, and we find the results are
satisfactory.

\section{Chemical Potential and Spin Distribution}   

In this section we will discuss the relation between quarks' chemical potentials and $\Delta u$, $\Delta d$,
 $\Delta s$, $R$. Many experiments show that we should take quarks with spin parallel to the nucleon's spin ($q\uparrow$)
different than those with spin anti-parallel ($q\downarrow$). The main compoents of the nucleon, except for the gluons,
are $u\uparrow$, $u\downarrow$, $d\uparrow$, $d\downarrow$, $s\uparrow$, $s\downarrow$ and their anti-particles
$\bar{u} \downarrow$, $\bar{u}\uparrow$, $\bar{d} \downarrow$, $\bar{d}\uparrow$, $\bar{s} \downarrow$, $\bar{s}\uparrow$ [7].
We assume that  heavy quarks $c$, $b$, $t$ almost do not appear [3].\\
In the nucleon, there are 12 constraints on the numbers and chemical potentials of the above 12 types of quarks. If $q$
stands for $u, d, s$ quarks generally,  $\Delta q$ stands for $\Delta u, \Delta d, \Delta s$, 
$n_q$ stands for valence quark number and ${\mu}_q$ stands for chemical potential,
we have the constraints as follow:
\begin{equation}
n_{q\uparrow}-n_{\bar{q}\downarrow}=\frac {(n_q+\Delta q)} {2};
\end{equation}
\begin{equation}
n_{q\downarrow}-n_{\bar{q}\uparrow}=\frac {(n_q-\Delta q)} {2};\\
\end{equation}
\begin{equation}
{\mu}_{q\uparrow}+{\mu}_{\bar{q}\downarrow}=0;\\ 
\end{equation}
\begin{equation}
{\mu}_{q\downarrow}+{\mu}_{\bar{q}\uparrow}=0. 
\end{equation}
By integrating eq (1) over x, we obtain that:
\begin{equation}  
n_q={\int}_{0}^{M/2} 3f({\mu}_q) \rho(E) dE
\end{equation}  
Using the density of states deduced in section 2 and assuming that the nucleon's mass is much biger than ${\mu}_q$, we get:
\begin{equation}  
n_q=\frac{3RT} {\pi} \ln [\exp(\frac {{\mu}_q}{T})+1]
\end{equation}  
From eq (7), (8), (9), (10), (12), we get:
\begin{equation}  
{\mu}_{q\uparrow}=\frac{(n_q+\Delta q)\pi} {6R},
\end{equation}  
\begin{equation}  
{\mu}_{q\downarrow}=\frac{(n_q-\Delta q)\pi} {6R}.
\end{equation}  
The relation between $\Delta q$ and chemical potentials is:
\begin{equation}  
\Delta q=\frac{3R({\mu}_{q\uparrow}-{\mu}_{q\downarrow})} {\pi}.
\end{equation}  
From several experimental data [10], [11], [12], we use $\Delta u=0.83$, $\Delta d=-0.43$, $\Delta s=-0.10$ for the proton
and $\Delta u=-0.40$, $\Delta d=0.86$, $\Delta s=-0.06$ for the neutron in the following sections. 
\\
Once the nucleon radius $R$ is determined, all the chemical potentials are known from eq (13), (14), (9), (10). 
Even when $R$ is unknown, we still can get three conclusions from those equations: \\
1. The ratios of each chemical potentials are:
 ${\mu}_{u\uparrow}$ : ${\mu}_{u\downarrow}$ : ${\mu}_{d\uparrow}$ : ${\mu}_{d\downarrow}$ : 
${\mu}_{s\uparrow}$ : ${\mu}_{s\downarrow}$= 2.83 : 1.17 : 0.57 : 1.43 : -0.10 : 0.10
for the proton and 
${\mu}_{u\uparrow}$ : ${\mu}_{u\downarrow}$ : ${\mu}_{d\uparrow}$ : ${\mu}_{d\downarrow}$ : 
${\mu}_{s\uparrow}$ : ${\mu}_{s\downarrow}$= 0.60 : 1.40 : 2.86 : 1.14 : -0.06  : 0.06
for the neutron. \\
2. For both proton and neutron, 
$u\uparrow$, $u\downarrow$, $d\uparrow$, $d\downarrow$, $\bar {s} \downarrow$, $s\downarrow$ have positive
 chemical potentials while their anti-particles have negative chemical potentials. \\
3. The absolute value of the chemical potentials of $s\uparrow$, $s\downarrow$, $\bar {s} \uparrow$, $\bar {s} \downarrow$ are much 
smaller than others.

\section{T-R-M Function of the Nucleon}    
In this section we will discuss the relation between nucleon's temperature, radius and mass and finally determine
the value of $R$ and $T$.\\
The longitudinal momentum fraction of each types of quarks could be obtained by integrating over x the eq (1) mutiplied x:
 \begin{equation}  
p_q=\frac{4}{3M} {\int}_{0}^{M/2} 3f({\mu}_q) \rho(E) E dE.
\end{equation}  
Similarly, the longitudinal momentum fraction of the gluons is:
\begin{equation}  
p_g=\frac{4}{3M} {\int}_{0}^{M/2} 16f_B(0) \rho(E) E dE,
\end{equation}  
where $f_B(0)$ is the Bose distribution function and the chemical potential of the gluons is 0.
Our calculation shows that for both proton and neutron, $p_{u \uparrow}$, $p_{u \downarrow}$, $p_{d \uparrow}$, 
$p_{d \downarrow}$ and $p_g$ are relatively large, while $p_{s \uparrow}$, $p_{s \downarrow}$, $p_{\bar {s} \uparrow}$, 
$p_{\bar {s} \downarrow}$ are relatively small and $p_{\bar {u} \uparrow}$, $p_{\bar {u} \downarrow}$, 
 $p_{\bar {d} \uparrow}$, $p_{\bar {d} \downarrow}$ are relatively tiny. 
From eq (16), (17), we have:
\begin{equation}
p_{u \uparrow}=\frac {\pi {(n_u+\Delta u)}_{}^{2}} {18MR} + \frac {2\pi{T}_{}^{2}R} {3M};
\end{equation} 
\begin{equation}
p_{u \downarrow}=\frac {\pi {(n_u-\Delta u)}_{}^{2}} {18MR} + \frac {2\pi{T}_{}^{2}R} {3M};
\end{equation} 
\begin{equation}
p_{d \uparrow}=\frac {\pi {(n_d+\Delta d)}_{}^{2}} {18MR} + \frac {2\pi{T}_{}^{2}R} {3M};
\end{equation} 
\begin{equation}
p_{d \downarrow}=\frac {\pi {(n_d-\Delta d)}_{}^{2}} {18MR} + \frac {2\pi{T}_{}^{2}R} {3M};
\end{equation} 
\begin{equation}
p_g=\frac {32\pi{T}_{}^{2}R} {9M};
\end{equation} 
\begin{equation}
p_{s \uparrow}=p_{s \downarrow}=p_{\bar {s} \uparrow}=p_{\bar {s} \downarrow}= 
 \frac {\pi{T}_{}^{2}R} {3M};
\end{equation} 
\begin{equation}
p_{\bar {u} \uparrow}=p_{\bar {u} \downarrow}=p_{\bar {d} \uparrow}=p_{\bar {d} \downarrow} \approx 0,
\end{equation}
where $n_u$ and $n_d$ are valence quark numbers.
Note that the sum of all longitudinal momentum fractions is 1, that is, 
\begin {equation}
p_{u \uparrow}+p_{u \downarrow}+p_{d \uparrow}+p_{d \downarrow}+p_g+p_{s \uparrow}+p_{s \downarrow}+
p_{\bar {s} \uparrow}+p_{\bar {s} \downarrow}+p_{\bar {u} \uparrow}+p_{\bar {u} \downarrow}+
p_{\bar {d} \uparrow}+p_{\bar {d} \downarrow}=1.  
\end{equation}
From eq (18), (19), (20), (21), (22), (23), (24), (25) we get the T-R-M function as follows:
\begin {equation}
M=\frac {68\pi T} {9} \left(RT+ \frac {5+{\Delta u}_{}^{2}+{\Delta d}_{}^{2} }{68RT}  \right).
\end{equation}
The total energy of the sphere $E_T$ includes the zero point Casimir energy $\frac{C}{R}$, where $C \approx 2 $:
\begin{equation}
E_T=M+\frac{C}{R}
\end{equation}
We determine the nucleon radius $R$ by the total energy minimal principle:
\begin {equation}
{\left (\frac {\partial E_T} {\partial R}\right )}_{T}{}=0,
\end{equation}
and therefore get: 
\begin {equation}
R=\frac {2\pi (5+{\Delta u}_{}^{2}+{\Delta d}_{}^{2}+\frac {9C}{2\pi}) }{9M},
\end{equation}
\begin {equation}
T=\frac {9M \sqrt{(5+{\Delta u}_{}^{2}+{\Delta d}_{}^{2}+\frac {9C}{\pi} )}}
 {4\pi \sqrt {17}( 5+{\Delta u}_{}^{2}+{\Delta d}_{}^{2}+\frac {9C}{2\pi} )},
\end{equation}
For the proton, if we use M=938 MeV, $\Delta u=0.83$, $\Delta d=-0.43$, $C=2$,  we get
$R={(154 MeV)}_{}^{-1}=1.28$ fm,
$T=63.5$ MeV.   And for the neutron, 
if we use M=939 MeV, $\Delta u=-0.40$, $\Delta d=0.86$, $C=2$, we get $R={(153 MeV)}_{}^{-1}=1.29$ fm,
$T=63.4$ MeV. All chemical potentials can be obtained from eq (13), (14), (9), (10).
 Both the proton and neutron have similar radius and temperature, although the proton has a slightly
smaller radius and higher temperature. We may take their radius and temperature the same in some
approximations.

\section{Quark Distribution Functions Near x=1}   
After we get the $R$, $T$ and chemical potentials of each type of the quarks, from eq (1), we can obtain the quark distribution
functions near x=1. We assume that x is so near 1 that $1-x << \frac {2T}{M} \approx 0.135$. From eq (1), 
we have
\begin{equation}
q(x)=\frac{3RM(1-x)}{\pi x} \exp{\left (\frac {{\mu}_q-Mx/2}{T} \right )}.   
\end{equation}
So that,
\[
u(x)={u}_{\uparrow}^{}(x)+{u}_{\downarrow}^{} (x)+{\bar u}_{\uparrow}^{}(x)+ {\bar u}_{\downarrow}^{} (x) 
\]
\begin{equation} 
=\frac{12RM(1-x)}{\pi x} \exp{\left (\frac {-Mx}{2T} \right )} \cosh {\frac{n_u \pi}{6RT}} \cosh {\frac{\Delta u \pi}{6RT}};   
\end{equation} 
\[
\Delta u(x)={u}_{\uparrow}^{}(x)-{u}_{\downarrow}^{} (x)+{\bar u}_{\uparrow}^{}(x)-{\bar u}_{\downarrow}^{} (x) 
\]

\begin{equation} 
=\frac{12RM(1-x)}{\pi x} \exp{\left (\frac {-Mx}{2T} \right )} \sinh {\frac{n_u \pi}{6RT}} \sinh {\frac{\Delta u \pi}{6RT}};   
\end{equation}
\[
d(x)={d}_{\uparrow}^{}(x)+{d}_{\downarrow}^{} (x)+{\bar d}_{\uparrow}^{}(x)+ {\bar d}_{\downarrow}^{} (x) 
\]
\begin{equation}
=\frac{12RM(1-x)}{\pi x} \exp{\left (\frac {-Mx}{2T} \right )} \cosh {\frac{n_d \pi}{6RT}} \cosh {\frac{\Delta d \pi}{6RT}};   
\end{equation} 
\[
\Delta d(x)={d}_{\uparrow}^{}(x)-{d}_{\downarrow}^{} (x)+{\bar d}_{\uparrow}^{}(x)-{\bar d}_{\downarrow}^{} (x) 
\]
\begin{equation}
=\frac{12RM(1-x)}{\pi x} \exp{\left (\frac {-Mx}{2T} \right )} \sinh {\frac{n_d \pi}{6RT}} \sinh {\frac{\Delta d \pi}{6RT}};   
\end{equation} 
\[
s(x)={s}_{\uparrow}^{}(x)+{s}_{\downarrow}^{} (x)+{\bar s}_{\uparrow}^{}(x)+ {\bar s}_{\downarrow}^{} (x) 
\]
\begin{equation}
=\frac{12RM(1-x)}{\pi x} \exp{\left (\frac {-Mx}{2T} \right )} \cosh {\frac{\Delta s \pi}{6RT}};   
\end{equation} 
\begin{equation}
\Delta s(x)={s}_{\uparrow}^{}(x)-{s}_{\downarrow}^{} (x)+{\bar s}_{\uparrow}^{}(x)-{\bar s}_{\downarrow}^{} (x) 
=0
\end{equation} 
Thus, from eq (32), (33), (34), (35), (36), (37),  we can calculate 
$\frac {\Delta u(x)}{u(x)}$, $\frac {\Delta d(x)}{d(x)}$, $\frac {\Delta s(x)}{s(x)}$,
$\frac {d(x)}{u(x)}$, $\frac {s(x)}{u(x)}$, when $x \rightarrow 1$.
 For the proton, if we use the $R$, $T$ in section 4, we have $\frac {\Delta u(x)}{u(x)}=0.774$, $\frac {\Delta d(x)}{d(x)}=-0.425$ ( quite
in agreement with the quark-diquark model result [13], [14] ),  $\frac {\Delta s(x)}{s(x)}=0$,
$\frac {d(x)}{u(x)}=0.216$, $\frac {s(x)}{u(x)}=0.098$, when $x \rightarrow 1$. 
And for the Neuton, if we use the $R$, $T$ in section 4, we have $\frac {\Delta u(x)}{u(x)}=-0.397$, $\frac {\Delta d(x)}{d(x)}=0.786$, 
$\frac {\Delta s(x)}{s(x)}=0$, 
$\frac {d(x)}{u(x)}=4.83$, $\frac {s(x)}{u(x)}=0.466$, when $x \rightarrow 1$. So that, when
$x \rightarrow 1$, ${F}_{2}^{n}/{F}_{2}^{p}=2.155 {u}_{}^{n}(x)/{u}_{}^{p}(x)=0.453$, which is quite in agreement with
the recent experimental data [15] and the pQCD result [16], [17]. 

\section{Conclusion}   
The analytic statistical method in this paper is limitted in the large x region. However, the advantage of this method is
that it shows the relation between the quarks' chemical potentials and their spin distibutions, as well as the 
relation between the nucleon 's temperture, radius and mass, the so called T-R-M function. There are three
input parameters, $\Delta u$, $\Delta d$, $\Delta s$, which cannot be determined without any dynamical 
approach beyond the statistical mechanics. Therefore the statistical method in this paper is only a connection
between the phenomena, but it is helpful for our understanding.


\begin{thebibliography}{999}
\bibitem{1} R. P. Bickerstaff, J. T. Londergan, {\em Phys. Rev. D} {\bf 42} (1990) 3621.
\bibitem{2} M. G. Mustafa, A. Ansari, IP/BBSR/95-47.

\bibitem{3} R. S. Bhalerao, {\em Phys. Lett. B\/} {\bf 380} (1996) 1.
  
\bibitem{4} F. Buccella et al., {\em Mod. Phys. Lett. A\/} {\bf 13} (1998) 441.
  

\bibitem{5} V. V. Flambaum et al., {\em Phys. Rev. E\/} {\bf 57} (1998) 4933.

\bibitem{6} H. Lin, hep-ph/0105050.
 
\bibitem{7} R. S. Bhalerao, {\em Phys. Lett. B\/} {\bf 476} (2000) 285.


\bibitem{8} P. A. M. Guichon, {\em Phys. Lett. B\/} {\bf 200} (1988) 235.

 
\bibitem{9} A. Chodos, R. L. Jaffe, K. Johnson and C. B. Thorn, {\em Phys. Rev. D } {\bf 10} (1974) 2599.

  
\bibitem{10} E142, P. Anthony et al., {\em Phys. Rev. D \/} {\bf 54} (1996) 6620. 

\bibitem{11} HERMES, K. Ackerstaff et al., {\em Phys. Lett. B\/} {\bf 404} (1997) 383.
\bibitem{12} SMC, B. Adeva et al., {\em Phys. Rev. D \/} {\bf 58} (1998) 112002. 

\bibitem{13} B. -Q. Ma, {\em Phys. Lett. B \/} {\bf 375} (1996) 320.

\bibitem{14} W. Melnitchouk, A. W. Thomas, {\em Phys. Lett. B \/} {\bf 377} (1996) 11.

\bibitem{15} U. K. Yang, A. Bodek, {\em Phys. Rev. Lett.\/} {\bf 82} (1999) 2467. 
\bibitem{16} G. R. Farrar,  D. R. Jackson, {\em Phys. Rev. Lett.\/} {\bf 35} (1975) 1416. 

\bibitem{17} S. J. Brodsky, M. Burkardt,  I. Schmidt, {\em Nucl. Phys. B} {\bf 441} (1995) 197.  

\end{thebibliography}
\end{document}